\documentclass{osa-article}

\journal{osac}

\usepackage{bm}

\articletype{Research Article}

\begin{document}

\title{Snapshot  hyperspectral imaging via  spectral basis multiplexing in Fourier domain}

\author{ Chao Deng,\authormark{1,2,3}  Xuemei Hu,\authormark{1,3} Jinli Suo,\authormark{1,*}   Yuanlong Zhang,\authormark{1} Zhili Zhang,\authormark{2}  and Qionghai Dai\authormark{1}}

\address{
\authormark{1}Department of Automation, Tsinghua University, Beijing, 100084, China\\
\authormark{2}High-Tech Institute of Xi'an, Xi'an, 710025, China\\
\authormark{3} These authors contribute equally to this work
}
\email{\authormark{*}jlsuo@tsinghua.edu.cn} 



\begin{abstract}
Hyperspectral imaging is an important tool having been applied in various fields, but still limited in observation of dynamic scenes.
In this paper, we propose a snapshot hyperspectral imaging technique which exploits both spectral and spatial sparsity of natural scenes.
Under the computational imaging scheme, we conduct spectral dimension reduction and spatial frequency truncation to the hyperspectral data cube and snapshot it in a low cost manner.
Specifically, we modulate the spectral variations by several broadband spectral filters, and then map these modulated images into different regions in the Fourier domain. The encoded image compressed in both spectral and spatial are finally collected by a monochrome detector.
Correspondingly, the reconstruction is essentially  a Fourier domain extraction and spectral dimensional back projection with low computational load.
This Fourier-spectral multiplexing in a 2D sensor simplifies both the encoding and decoding process,  and makes hyperspectral data captured in a low cost manner.
We demonstrate the high performance of our method by quantitative evaluation on simulation data and build a prototype system experimentally for further validation.
\end{abstract}

\section{Introduction}
Hyperspectral imaging aims to capture the details of the spectra of natural scene . It is playing great roles in both scientific research and engineering applications, such as military security \cite{eismann1996comparison,ardouin2007demonstration}, environmental monitoring \cite{delalieux2009hyperspectral}, biological science \cite{backman2000detection,zavattini2006hyperspectral}, medical  diagnosis \cite{kester2011real,vo2004hyperspectral}, scientific observation \cite{barnsley2004proba,bibring2005mars}, and many other fields\cite{pan2003face,brady2009optical,arce2014compressive}.

Since only 1D  and 2D commercial imaging sensor are available, primary hyperspectral imaging are implemented in a scanning mode, either along spatial or spectral \cite{goetz1985imaging,sellar2005comparison} dimension. Although the advanced pixel count and sensitivity of array detectors boost the spatial and spectral resolution of scanning-based hyperspectral imaging systems, the requirement of steady scanning limits the speed and robustness of scanning based methods. Hence, there is a general scientific interest in investigating spectral imaging of dynamic samples such as combustion processes \cite{reiter2011combustion} or fluorescent probes used in biological and biomedical imaging \cite{kauranen1991spatial,volin1998high}, and snapshot hyperspectral imaging is one of the main research focuses in this area.

The typical way to perform snapshot hyperspectral imaging is mapping different spectral bands (either single narrow band or multiplexed ones) to different positions and then collecting them by one or more detectors. So far, various methods have been proposed to implement such band mapping.
The integral field spectroscopy (IFS) \cite{weitzel19963d,hagen2013review},  multispectral beamsplitting (MSBS) \cite{matchett2007volume},  image replicating image spectrometer (IRIS) \cite{gorman2010generalization,harvey2003high} and multispectral sagnac interferometer (MSI) \cite{kudenov2010white} are the paradigms directly using stacked optical components to split spectral channels, which complicate the system and impose strict limitation on the light path building. Another drawback of the direct mapping method is limited spectral band number. For example, the spectral band number of MSBS is dependent on the beam splitter which is upper limited by 4, and the spectral resolution of IRIS is limited to 16 wavelength bands due to lack of large-format  sufficient-birefringence Wollaston polarizers. In comparison, computed tomography imaging spectrometer (CTIS) \cite{okamoto1991simultaneous}, multi-aperture filtered camera  (MAFC) \cite{hirai1994application}, image mapping spectrometry (IMS) \cite{gao2009compact} and snapshot hyperspectral imaging Fourier transform (SHIFT) \cite{kudenov2010compact} are more compact but need customized components, which is usually expensive and of high fabrication difficulty. Introducing new reconstruction algorithms might inspire new imaging schemes and raise the final performance. The coded aperture snapshot spectral imager (CASSI)\cite{gehm2007single} is the first imager exploiting the compressive sensing theory to recover hyperspectrum.  Getting rid of simple combination of optical elements, this method demands careful calibration and heavy computational load, which limits its broad applications in the scenarios where online reconstruction is needed. In sum, a snap shot spectral imager, with simple implementation, low computation load, and high reconstruction performance are worth studying.

One inspiring work in this direction is the frequency recognition algorithm for multiple exposures (FRAME) \cite{dorozynska2017implementation,kristensson2017instantaneous,ehn2017frame}, which uses multiplexed sinusoidal illumination to encode several images into a single one simultaneously.
Making use of both the spatial \cite{chakrabarti2011statistics,bian2016efficient} and spectral sparsity of the natural hyperspectral data,
we propose a snapshot  Fourier-Spectral-Multiplexing (FSM) method for hyperspectral imaging using a monochrome camera.
 Dividing the image sensor into a few subfields is a simple way to record various channels simultaneously. This method either shrinks the field of view (FOV) or linearly decrease the pixel resolution. Differently,  considering that the  Fourier spectrum of nature scenes concentrate at the central low frequency region, FSM can effectively eliminate severe resolution loss and FOV shrinkage. 
Our snapshot hyperspectral imaging method can achieve high spatial-spectral resolution and high frame rate, with low cost  setup and computational workload. The qualitative performance comparison of the different state-of-the-arts is shown in Table \ref{table1}. 
\begin{table}
\centering
\caption{\textbf{The performance comparison of different hyperspectral imaging methods}}
\label{table1}
\begin{tabular}{ccccc}
  \hline
   & Spatial Res. & Spectral Res.  & Computational Load & Cost\\
  \hline
  IFS & low & low   & low & high \\
  MSBS & high & low  &  low & high \\
  IRIS & low & low  &  low & high \\
  MSI & high &  low  &  low& high  \\
  CTIS &low & low    &  low & high  \\
  MAFC & low &high  &  low & high \\
  IMS  & low  & high   &  low & high \\
  SHIFT &low& high   &  low & high \\
  CASSI& high & high  &  high & low   \\
  FRAME& high  & low &  low  & low       \\
  \textbf{FSM}  & \textbf{high} &\textbf{high} & \textbf{low} & \textbf{low}\\
  \hline
\end{tabular}
\end{table}
In summary, we make the following contributions: 

\begin{itemize}
	\item We \emph{combine} spectral dimension reduction and Fourier-Spectral-Multiplexing (FSM) strategy as a new way of low cost hyperspectral imaging.
	\item We \emph {validate} our method in simulation both quantitatively and qualitatively.
	\item We \emph{build} a snapshot hyperspectral imaging prototype to verify this approach and \emph{demonstrate} its practical benefits.
\end{itemize}

The structure overview of our paper is given as below. In section 2, we introduce our imaging scheme and formulation. In section 3, we set the parameters and  quantitatively evaluate the effectiveness in simulation. In section 4, we demonstrate our method with physical experiments. In section 5, we discuss the advantages and limitations of our method, and the future work of our imaging system.

\section{Method}

\subsection{Imaging scheme}

\begin{figure}[htb]
  \centering
  \includegraphics[width=1.0\linewidth]{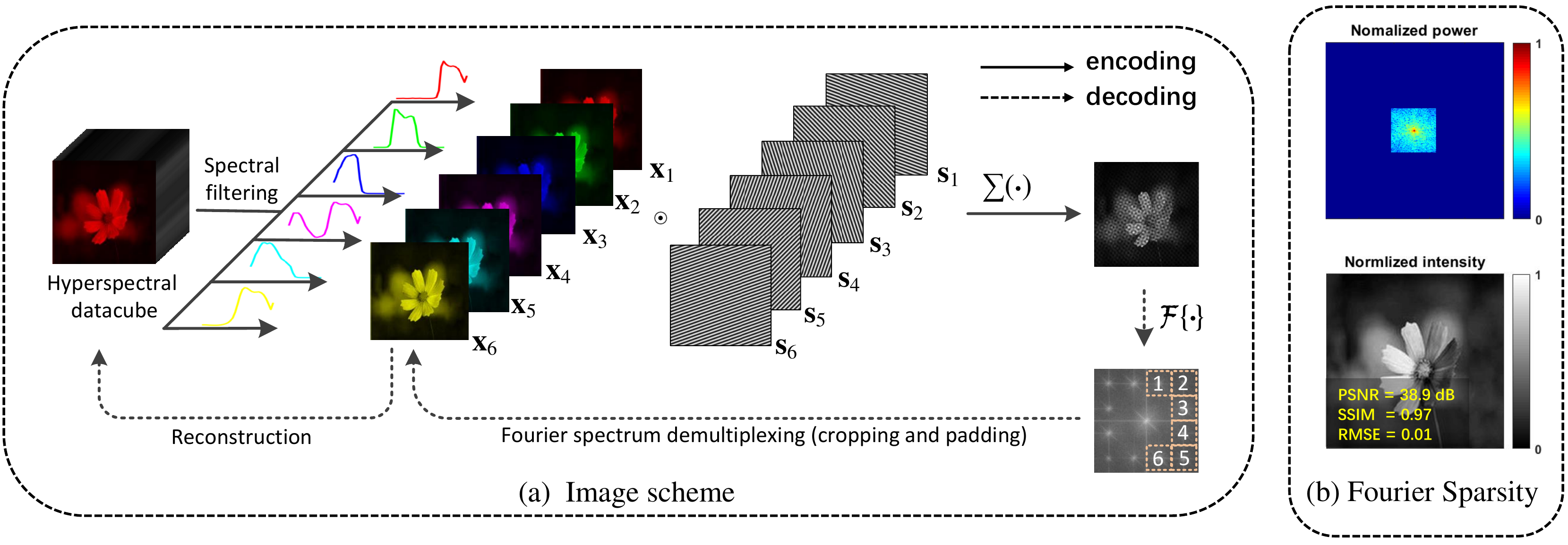}\\
  \caption{(a) The scheme of the proposed hyperspectral imaging system. The hyperspectral data is spectrally filtered and projected  into six images: ${\bf x}_1, {\bf x}_2, \cdots, {\bf x}_6$, and then codified by six sinusoidal patterns ${\bf s}_1, {\bf s}_2, \cdots, {\bf s}_6$, respectively.  The six sinusoidal patterns shift the Fourier distribution of six projected images away from the origin into six different regions, and the gray scale camera captures the modulated images in an add-up way. The hyperspectral images are reconstructed through a two-stage method, including: Fourier
spectrum demultiplexing and linear system reconstruction. (b) The centralized spreads of the nature images' Fourier coefficients. The reconstruction (bottom image) from 6.25$\%$ ($0.25^2$) Fourier coefficients locating at the centroid region (upper image) is quite clear. }\label{imagingprocess}
\end{figure}

The architecture for the snapshot hyperspecral imaging is shown in Fig.~\ref{imagingprocess}(a), including both encoding and decoding modules. During encoding, the 3D hyperspectral data cube is sequentially projected to a low dimensional space after going through a series of wide-band spectral filters, denoting as  ${\bf x}_1,{\bf x}_2, \cdots, {\bf x}_J$.
Since the spectrum of natural scenes are of low intrinsic dimension, the hyperspectral datacube can be denoted as the linear summation of a few spectral basis. To ensure that the reconstruction problem is well posed, the  number of color filters is equal to that of spectral basis.
 Statistics tells that six bases are sufficient for high fidelity  hyperspectral data representation (see \cite{yasuma2008generalized}), so here we set $J$ as 6.
 Then the projections are modulated by different sinusoidal patterns ${\bf s}_1,{\bf s}_2,\cdots, {\bf s}_J$, respectively. Mathematically, the sinusoidal modulation is represented as point-wise product ``$\odot$'', and would shift the Fourier spectrum of each projection into a specific region in the Fourier domain. The sinusoidal patterns are designed to ensure few overlap among the Fourier distributions of $\{{\bf x}_i\}$. Finally, the fast coded image is recorded by a gray scale camera in an add-up way as shown in the Fig.~\ref{imagingprocess}(a). Corresponding to the coding procedure, the  decoding   is straightforward.  We first transform the captured image to the Fourier domain, and extracting the Fourier spectrum of each projection ${\bf x}_i$ according to the shifting effect of the corresponding sinusoidal modulation ${\bf s}_i$.  Then we transform the separated Fourier spectrum back to the spatial domain. Finally, we back project the reconstructions to the high dimensional data cube by solving a linear system. As Fig.~\ref{imagingprocess}(b) shows, the Fourier coefficients of natural scenes mainly concentrate at the low frequency region, and the reconstruction from a small proportion of the center coefficients (padding zeros for the surrounding ones) can reconstruct the image with high quality. Therefore, we propose to encode multiple spectral projections into a single image in the Fourier domain.
 Correspondingly, the Fourier spectrum demultiplexing includes two processes: Fourier spectrum cropping and zero padding to the original resolution. In Sec.~\ref{subsec:formulation}, we introduce the formulation of the whole encoding and decoding process in details.
 
\subsection{Formulation}
\label{subsec:formulation}

Though going through the $i$th spectral filter,  the spectrum of the target scene is filtered with the corresponding spectral transmission, which can be represented as
 \begin{equation}\label{collectingmodel}
 {\bf x}_i=\int_\lambda {\bf I}_i(\lambda) {\bf r}(\lambda)d\lambda,
\end{equation}
in which  ${\bf I}_i(\lambda)$ is the spectrum transmission  of the $i$th spectral filter and ${\bf r}(\lambda)$ denoting the spectrum reflectance/transmission. Here we omit 2D spatial coordinate for  brevity.
Statistically, the spectra of the natural materials can be represented as a linear summation of a few (e.g., $J$=6) characteristic spectral basis \cite{parkkinen1989characteristic}, i.e.
\begin{equation}\label{hyperbasis}
{\bf r}(\lambda)=\sum_{j=1}^J {\bm \alpha}_j{\bf b}_j(\lambda),
\end{equation}
where ${\bf b}_j$ is the $j$th spectral basis, and ${\bm \alpha}_j$ is the corresponding coefficients.
Substituting Eq.~\eqref{hyperbasis} into Eq.~\eqref{collectingmodel} \cite{park2007multispectral,han2010fast}, we could get
\begin{equation}\label{hyperreconstruction}
{\bf x}_i=\sum_{j=1}^J {\bm \alpha}_j\int_\lambda {\bf I}_i(\lambda){\bf b}_j(\lambda) d\lambda.
\end{equation}
In Eq. \eqref{hyperreconstruction}, ${\bf I}_i(\lambda)$ is precalibrated and ${\bf b}_j(\lambda)$ is trained from hyperspectral database \cite{parkkinen1989characteristic}. So the hyperspectrum of the surface can be represented by $J$ parameters ${\bm \alpha}_j$.

The $J$ sinusoidal patterns $\{{\bf s}_i\}$ for Fourier domain multiplexing of $J$ projected images $\{{\bf x}_i\}$ are designed to have as minimal overlap in the Fourier domain as possible.  Generally, the sinusoidal pattern can be written as
\begin{equation}\label{sinePattern}
{\bf s}_i=1+\cos({\bf p}\cdot {\bm \omega}_i),
\end{equation}
where ${\bf p}$ is the 2D spatial coordinate, ${\bm \omega}_i$ ($i=1,2,\cdots,J$) represents the spatial frequency, and the Fourier transform of Eq.~\eqref{sinePattern} includes three delta functions $\delta({\bm \omega}+{\bm \omega}_i)$, $\delta({\bm \omega})$, $\delta({\bm \omega}-{\bm \omega}_i)$ in the Fourier domain.
After applying such sinusoidal modulation to the target scene, its spatial spectrum is duplicated into three replicas centered at  ${\bm \omega}=-{\bm \omega}_i,\;{\bf 0},\;{\bm \omega}_i$, respectively. Research in the field of natural image statistics suggests that the Fourier domain of natural images is sparse \cite{rabbani2002jpeg2000} and enables efficient coding and sampling of natural images \cite{bian2016efficient}. Taking advantage of this prior information, if the shifted distance $\|{\bm \omega}_i\|$ is set properly, these three replicas can be separated to eliminate aliasing effectively.
To set the sinusoidal patterns statistically, we first analyze the Fourier spectrum of natural scenes and then estimate the proper shift distance and the corresponding cropping size accordingly (details are discussed in Sec.~\ref{sec:parametersetting}).

The encoded image is recorded by the gray scale camera as
\begin{equation}\label{record}
{\bf y}=\sum_{i=1}^J{\bf x}_i\odot{\bf s}_i.
\end{equation}
In terms of reconstruction, we first transform the captured image ${\bf y}$ to Fourier domain and demultiplexing each image by Fourier spectrum cropping and padding according to each sinusoidal pattern. After inverse Fourier transform to spatial domain, we obtain estimations of $J$ spectral filtered images $\{\hat{{\bf x}}_i\}$.  To further remove the aliasing, we use generalized alternating projection (GAP)  to improve the reconstruction using these estimations as the initial value. GAP is basically an extended  alternating projection algorithm which solves the compressive sensing problem in the transformed domain, i.e.  discrete cosine transformation or wavelet domain \cite{liao2014generalized,yuan2016generalized}. The optimization problem is formulated as
\begin{equation} \label{GAP}
\min_{\bf W} \|{\bf W}\|_{\ell_{2,1}^{\mathcal{G}\  \beta}}, \quad \text{subject to }\; {\bf Y}={\bf S}{\bf X} \quad \text{and} \quad {\bf X}={\bf TW},
\end{equation}
where the capital notation  ${\bf Y}$ and ${\bf X}$ are  vectorized forms of ${\bf y}$ and ${\bf x}$ [e.g., ${\bf Y} = \text{vec}({\bf y})$, ${\bf X} = \text{vec}({\bf x})$], respectively.  ${\bf S}=[{\bf S}_1,\ {\bf S}_2,\ \cdots,\  {\bf S}_J]$  with ${\bf S}_i$ being diagonalization of vectorized form of ${\bf s}_i$ [e.g., ${\bf S}_i = \text{diag}(\text{vec}({\bf s}_i))$].  ${\bf T}$ is wavelet transformation matrix, and ${\bf W}$ is corresponding coefficient in the transformed domain. 

 The $\|\cdot\|_{\ell_{2,1}^{\mathcal{G}\  \beta}}$ is weighted group-$\ell_{2,1}$ norm, calculated as
\begin{equation}
\|{\bf W}\|_{\ell_{2,1}^{\mathcal{G}\  \beta}}=\sum_{k=1}^m\beta_k\|{\bf W}_{\mathcal{G}_k}\|_2,
\end{equation}
where  $m$ is group number, ${\bf W}_{\mathcal{G}_k}$ is a subvector of ${\bf W}$ from $k$th group,  with $\mathcal{G}_k$ being components  index and $\beta_k$ being the group weight coefficient.  The Eq.~\eqref{GAP} would converge with desired accuracy after a number of iterations. After de-aliasing, we obtain estimations of six spectral-filtered images $\{\hat{{\bf x}}_i\}$ with high quality.

To obtain the hyperspectral images, we inverse the linear system defined in Eq. \eqref{hyperreconstruction} and get the following constrained optimization:
\begin{equation}
\arg\min_{\bm \alpha} \|\hat{{\bf x}}-{\bf F}{\bm \alpha}\|_2^2+\eta\big\|\frac{\partial^2 {\bf r}(\lambda)}{\partial\lambda^2}\big\|_2^2\quad \text{s.t.}\quad {\bf r}(\lambda)\geq0.
\label{quadraticProgammning}
\end{equation}
Here  $\hat{{\bf x}} = [\hat{{\bf x}}_1,\  \hat{{\bf x}}_2,\  \cdots,\  \hat{{\bf x}}_J]^T$, ${\bm \alpha}=[{\bm \alpha}_1,\ {\bm \alpha}_2,\ \cdots,\ {\bm \alpha}_J]^T$, and ${\bf F}_{ij}=\int_\lambda {\bf I}_i(\lambda){\bf b}_j(\lambda)d\lambda$. The parameter $\eta$ weights the spectrum smoothness and we set $\eta=200$ in implementation empirically. This constrained optimization problem  is solved with the  quadratic programming solver in Matlab, which is based on the Lagrangian multiplier method. 

\section{Parameter setting and quantitative evaluation}
\label{sec:parametersetting}

\begin{figure}[htb]
  \centering
  \includegraphics[width=1.0\linewidth]{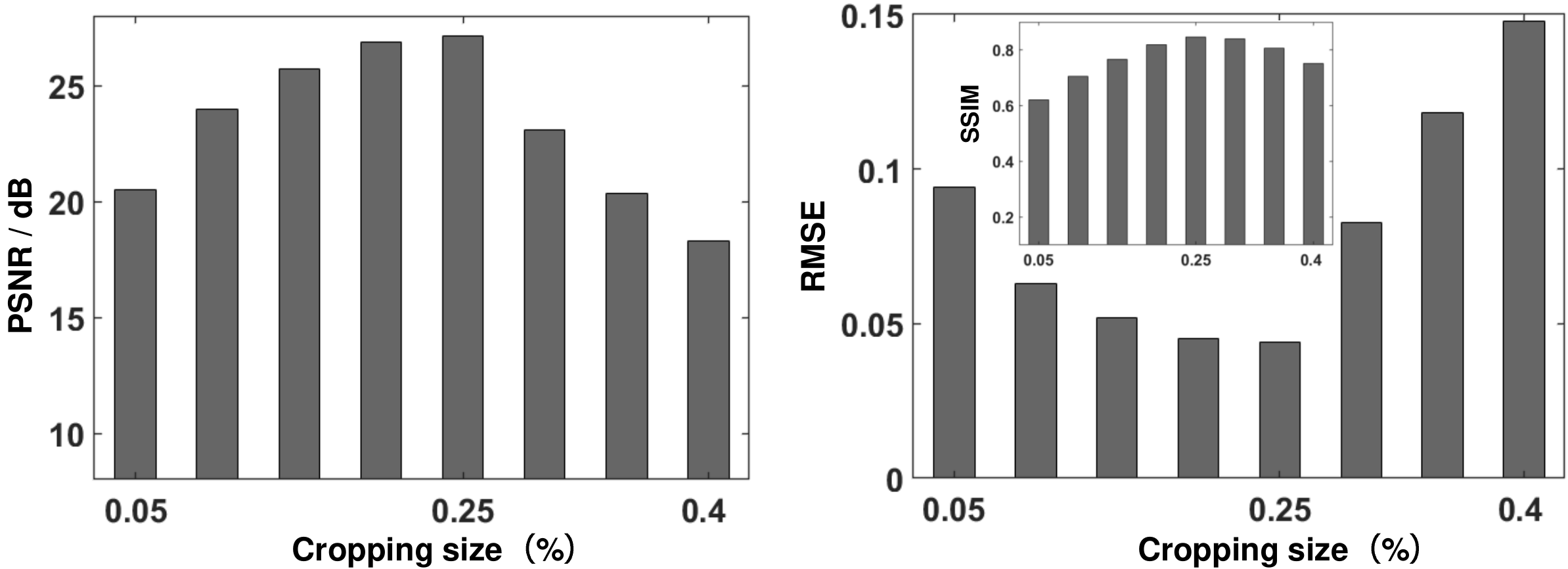}\\
  \caption{The PSNR, RMSE and SSIM  scores of different cropping sizes, ranging from 0.05 to 0.4 of image width.}\label{croppingsize}
\end{figure}

The  cropping size is designed from following aspects: (i) The Fourier distributions of six filtered versions are slightly different but similar, so we set the cropping size uniformly; (ii) To minimize the crosstalk, the six modulation frequencies should be located away from the origin and adjacent version; (iii) The cropped regions should occupy as much area in the Fourier domain to avoid losing many details in the final reconstruction. To compare different cropping sizes, we simulate the performance in terms of  peak signal-to-noise ratio (PSNR),  root mean square error (RMSE) and structural similarity (SSIM)\cite{wang2004image,dosselmann2011comprehensive,wang2011information}, the result is shown in Fig.~\ref{croppingsize}. Specifically, the cropping size ranges from 0.05 to 0.4 of image width, and the PSNR, RSME and SSIM  values are the average of six Fourier de-multiplexing channels, respectively. From the simulation, we can see that the optimum performance is achieved when the cropping size is set to 0.25 of image width.
Moreover, we conduct a statistical analysis over the multi-spectral database built by the CAVE laboratory at Columbia University \cite{yasuma2010generalized} to evaluate the 0.25 Fourier cropping, in terms of PSNR, RMSE and SSIM, these three metrics maintain at 32.0 dB, 0.03 and 0.94 on average.
We use the same modulation scheme for different samples in order to develop a general imager. The scheme is applicable for most nature scenes, because their spectral spreads are similar and applying GAP de-aliasing can handle the slight variance among samples.

 \begin{figure}[htb]
  \centering
  \includegraphics[width=0.7\linewidth]{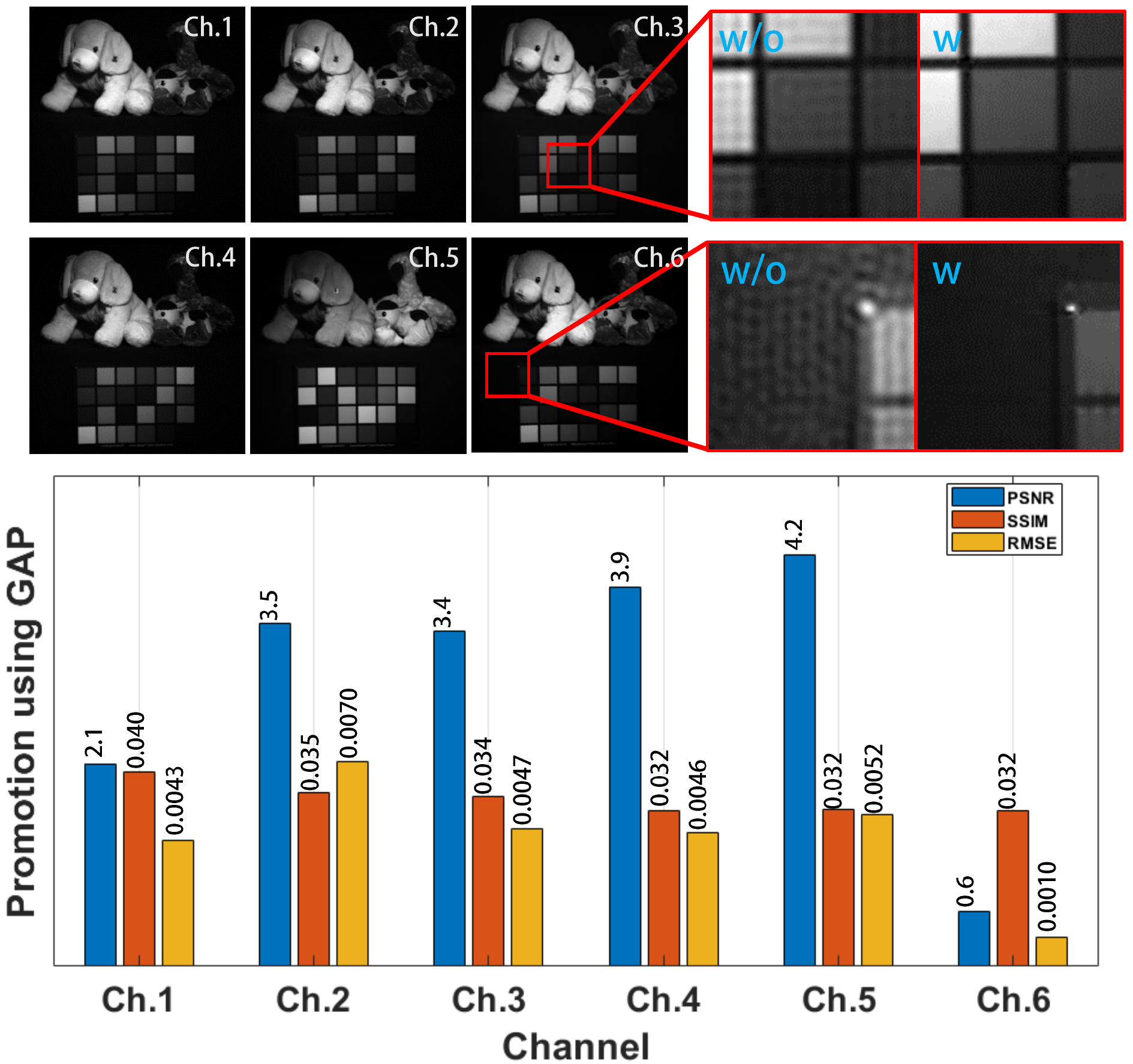}\\
  \caption{Performance promotion by GAP. \textbf{Upper part}: six channels reconstructed after GAP optimization. Ch.1$\sim$Ch.6 represent the six projections, and the close-ups compare the de-aliasing before and after using GAP. w/o: without GAP optimization;  w: with GAP optimization. \textbf{Bottom part}: PSNR, SSIM and RMSE \textit{promotion} through GAP optimization.}\label{simulation_optimization}
\end{figure}

During de-multiplexing of the spatial projections from the recorded encoded measurement, as aforementioned, we use GAP algorithm after Fourier extracting  to further suppress aliasing.
The simulation on the CAVE multi-spectrum database reveals 3.05 dB improvement on average by introducing GAP optimization. Fig.~\ref{simulation_optimization} displays the result on an example image with and without GAP optimization. We can clearly see the improvement introduced by GAP, in terms of PSNR, RMSE, and SSIM.

\begin{figure}[htb]
  \centering
  \includegraphics[width=0.8\linewidth]{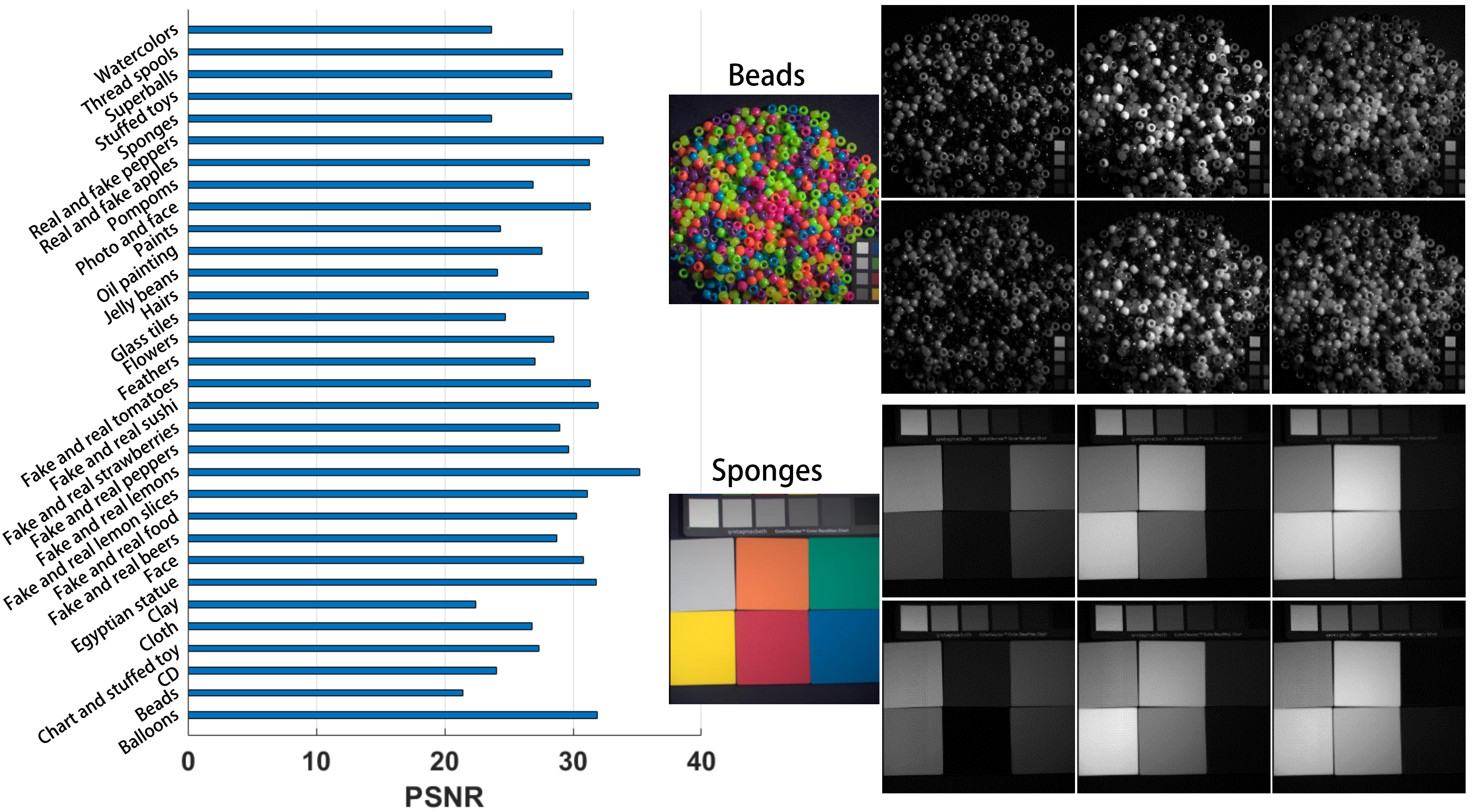}\\
  \caption{The PSNR of the reconstruction and two examples from the CAVE multi-spectral database. In each example, the upper row is the ground truth in 500 nm, 600 nm, and 700 nm, respectively and the lower row is the corresponding reconstruction. }\label{simulationresult}
\end{figure}

To evaluate the performance of our hyperspectral imaging system quantitatively, we test the imaging accuracy on the CAVE's multi-spectral image database. For each example, we simulate the encoded image based on the  six sinusoidal patterns $\{{\bf s}_i\}$ and CW's spectral responses $\{{\bf I}_i\}$ demonstrated in Fig.~\ref{setup}(c). The reconstruction results are quite promising, with PSNR averagely over 28.4 dB, as shown in the left subfigure of Fig.~\ref{simulationresult}. For a clearer demonstration, we also display two scenes with the lowest PSNR in the right part: Beads and Sponges, each with the ground truth (upper row) and corresponding reconstruction (lower row) displayed in comparison. We can see high fidelity reconstruction are achieved.

\section{Imaging prototype and experiment on captured data}
\begin{figure}[htb]
  \centering
  \includegraphics[width=0.85\linewidth]{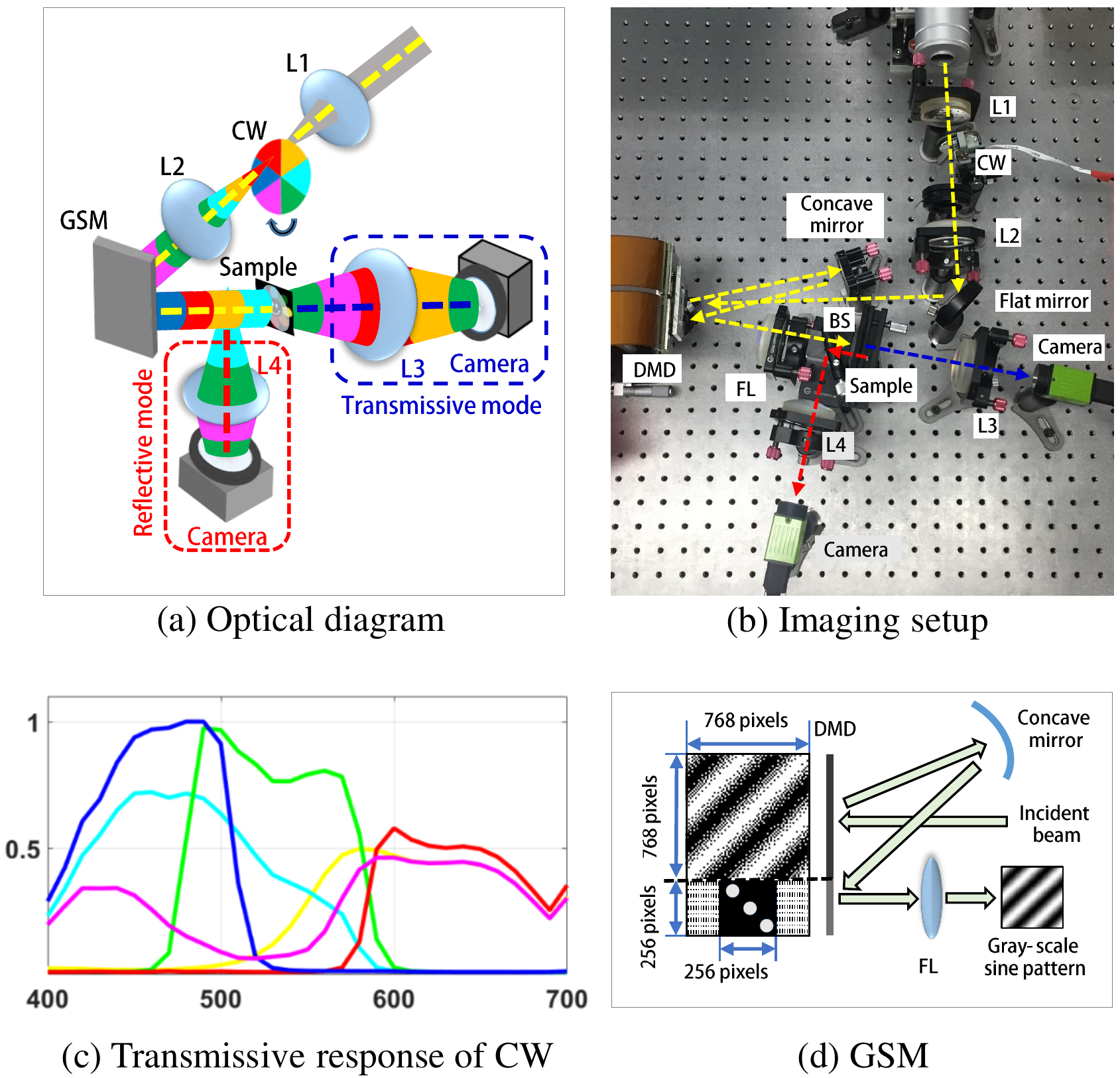}\\
  \caption{The imaging scheme of proposed method. (a) The optical diagram including transmissive and reflective mode (the beam splitter is omitted in reflective mode). L1 and L2: converging lens composing a 4f system. CW: color wheel.  GSM: gray scale sinusoidal modulation. L3 and L4: Converging lens. (b) The imaging setup including both transmissive and reflective mode. DMD: digital micromirror device; FL: Fourier lens.  (c) The transmissive response of CW. (d) The GSM module implementing fast gray scale sinusoidal modulation. }\label{setup}
\end{figure}

\begin{figure}[htb]
  \centering
  \includegraphics[width=0.85\linewidth]{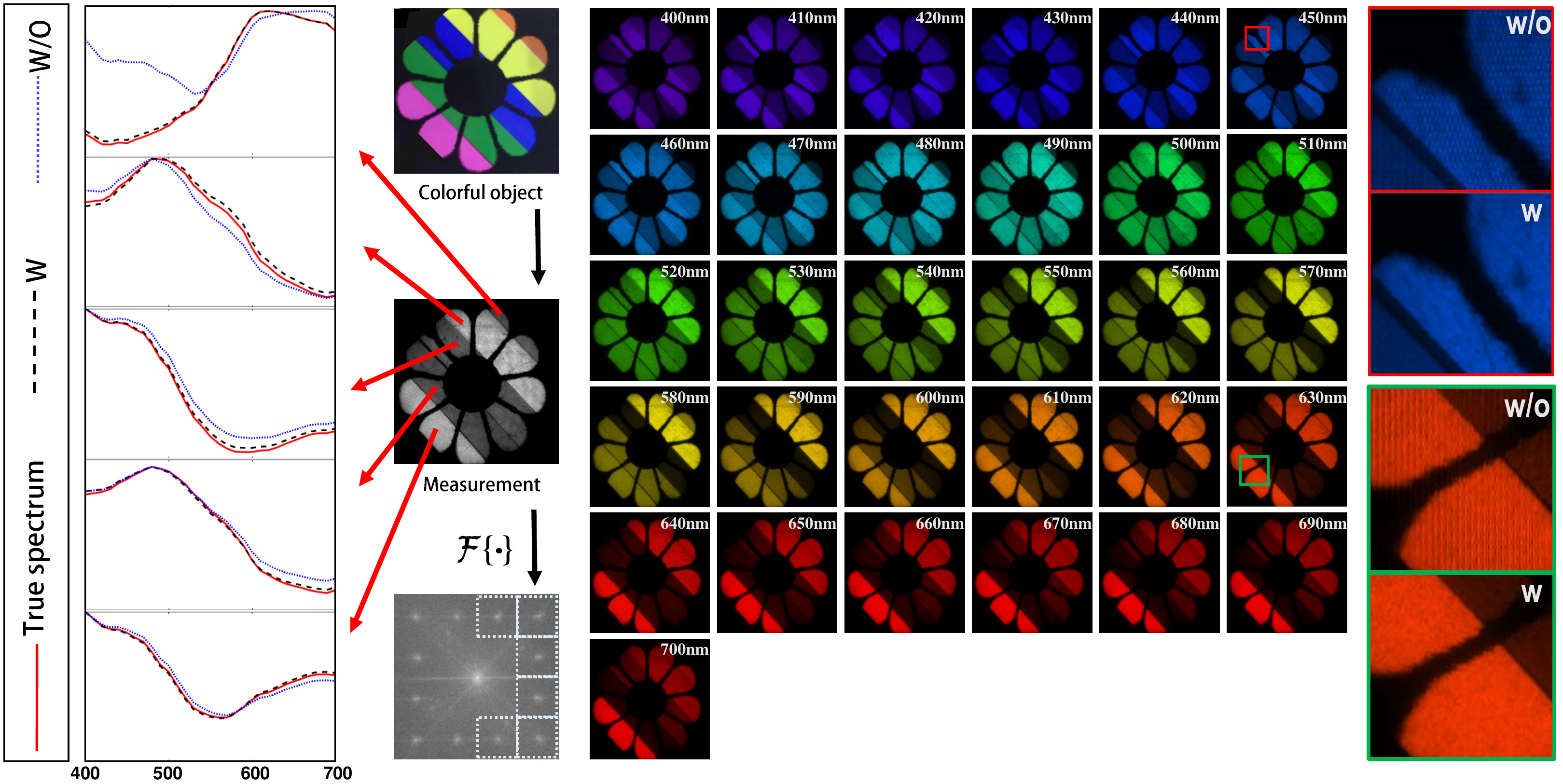}\\
  \caption{The experimental reconstruction of  a flower film with five strips of different transparent color paper. The RGB image of the object,  the measurement together with its Fourier space and the reconstructed five spectrum of  transparent color paper  are displayed in the left part. The solid red,  dashed black and dotted blue curve are the true spectrum, reconstruction with (w) GAP and without (w/o) GAP, respectively. The  reconstructed  hyperspectral images are displayed in the middle. The spectral range is 400 nm $\sim$ 700 nm. We highlight the distinction of reconstruction without  and with GAP optimization in the right part.}\label{reconstruction1}
\end{figure}

\begin{figure}[htb]
  \centering
  \includegraphics[width=0.85\linewidth]{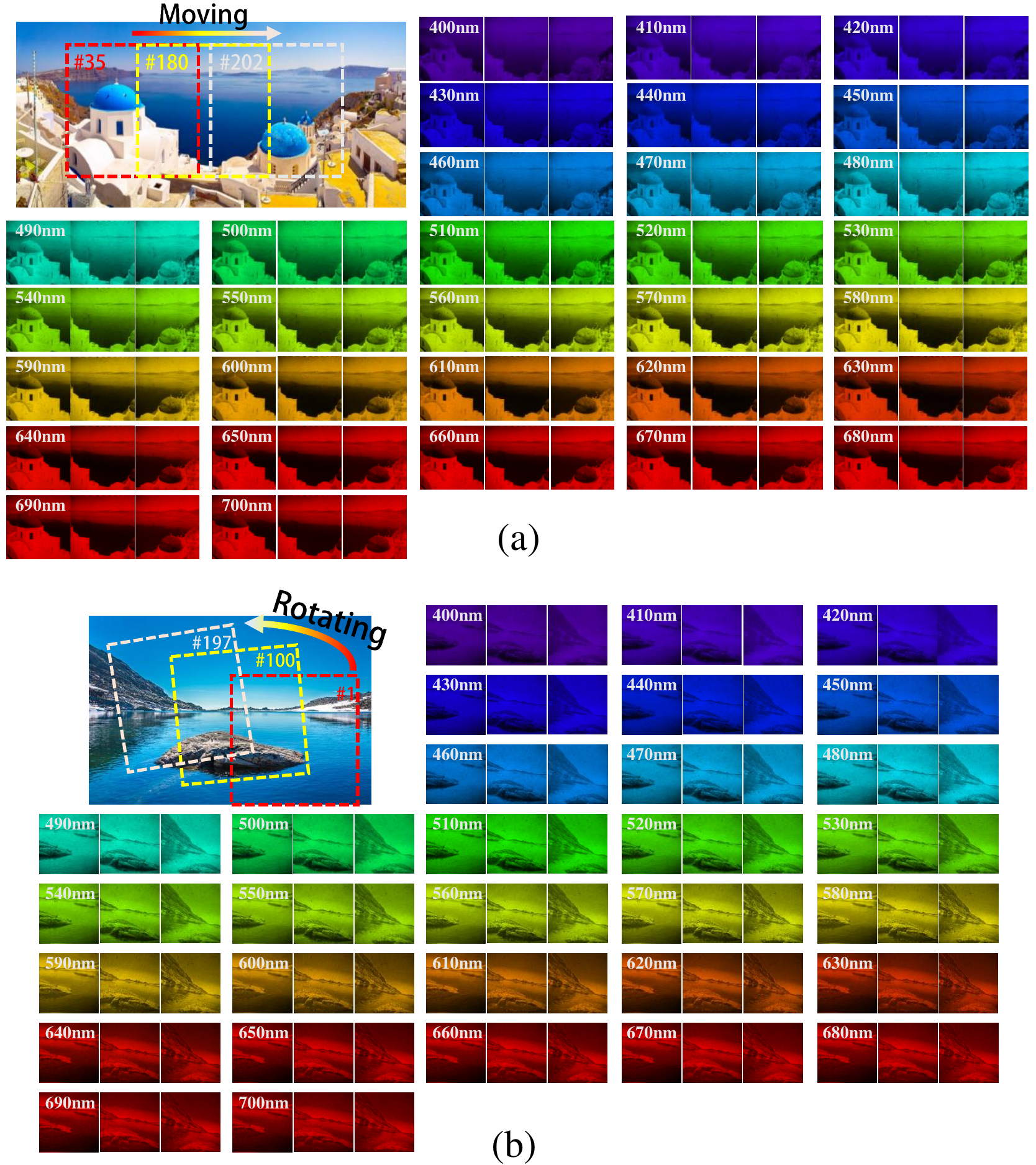}\\
  \caption{Hyperspectral reconstruction of regular motion: (a):  a color film of Santorini Island  mounted on a translation stage moving at a constant speed of 3.7 mm/s; (b): a color film of a landscape fixed on a rotary translation stage with angular speed of 0.023 rad/s. }\label{reconstructionvedio}
\end{figure}

We build a prototype to verify our imaging scheme as shown in Fig.~\ref{setup}.  The collimated broadband light is first modulated by a synchronized rotating color wheel (CW).  The  off-the-shelf six segment color wheel is driven by a high speed motor (a 60k rpm Walkera Super CP motor) and placed on the Fourier plane of the 4f system composed of L1 and L2 for spectral modulation. The spot on the color wheel is small enough so that the  large temporary overlap between two segments can be neglected.
 As the correlation among the spectral responses of the off-the shelf segments is high, we conduct a slight modification. Specifically, we choose three segments (red, green and blue) from the off-the-shelf color wheel, and attach three other broadband color papers to the transparent segment. The three new spectral filters are selected according to two criteria. First, for effective reconstruction, the spectrum of these filters should cover the target spectrum. Second, from an available spectral filters set, we traverse all the possible combinations to choose the optimum filter group with minimum correlation by following optimization
\begin{equation}
\min \max_{\substack{1\leq i,j\leq 6,\\ i\neq j}} corr(I_i,I_j),
\end{equation}
in which $I_i, I_j$ are two spectral filters in each group, and the notation $corr$ represents correlation of two vectors. Except for large ratio between bandwidth and the number of reconstructed wavelength \cite{wang2014computational}, broadband spectral filters has higher signal-to-noise ratio than narrowband ones.
 Then a gray scale sinusoidal pattern module (GSM) modulates the incoming light .
Specifically, we use a DMD to modulate the binary approximation of gray scale sinusoidal patterns to achieve full modulation rate, and apply pinhole filtering on the Fourier plane to obtain ideal sinusoidal patterns. As for binarization, we choose dithering algorithm considering its low approximation error.
 In implementation, we use a single DMD to conduct binary patterning and Fourier-domain filtering simultaneously. The left part of the DMD modulates the incident light beam with a dithering sinusoidal pattern, and  the right half of the DMD optically filters the spatial spectrum to remove the approximation error introduced by dithering algorithm [Fig.~\ref{setup}(d)]. The filter selects three dominant frequencies of an ideal sinusoidal pattern, i.e., -1, 0, and +1. Here the Fourier transform is taken by a concave mirror, which is well designed so that the right region of DMD is the Fourier plane of the left counterpart. This modulation can achieve 20 kHz gray scale sinusoidal patterning and is sufficient for our imaging scheme.
To properly configure the optical elements in the narrow space, we slightly rotate the DMD plane so that its incident and outgoing beams are of a small angle. Moreover, to achieve compact optical design, we can also exploit short-focus lenses with small aperture and customized optical bracket. We use the GSM modulation for three reasons. First, unlike ideal gray scale sinusoidal modulation, digital representation could introduce unwanted frequency components in the Fourier domain and thus deteriorate the final reconstruction. Second, the GSM arrangement can also block the diffraction orders introduced by DMD. Third, as the refresh rate of gray scale DMD modulation is upper limited to 253 Hz (8 bit), we adopt the same method proposed in  \cite{zhang2018doubling} to implement fast sinusoidal modulation for scalability, with the scheme illustrated in Fig.~\ref{setup}(d).
After modulation, the light beam is focused by the converging lens onto the sample.   Finally, the encoded sample information is collected by a gray scale camera (GO-5000C-USB, JAI) after converged by a lens.   

In our experiment,  the pixel resolution of  both  sinusoidal patterns displayed in DMD and the detector are $768\times768$, and  the Fourier cropping size is 192 pixels ($768\times 25.0\%$).
The spectral resolution is mainly determined by three factors: First, the number of PCA bases, as mentioned in Chapter 4. The spectral resolution increases with the number of PCA bases (e.g., eight in \cite{parkkinen1989characteristic}). Second, the selection of the transmissive spectrums of the color wheel segments. We can obtain higher spectral resolution by reducing the correlation among the transmissive spectrums. Third, the regularization parameter. An improper regularization parameter $\eta$ in Eq.~\ref{quadraticProgammning} would degenerate the spectral resolution: a too small $\eta$ cannot suppress noise effectively, while a too large $\eta$ would smooth out the spectral curve. Empirically, we set $\eta=200$ in implementation.
 We use external trigger mode for synchronization: the trigger out signal per circle of the color wheel  triggers  the DMD for displaying the sinusoidal patterns, and again the trigger out signal of DMD triggers  the exposure of the camera.
The final imaging speed is limited by the lowest speed of three elements -- the rotation of color wheel (1000 Hz), camera's frame rate ($\sim100$ Hz), and DMD refresh rate (20000 Hz).
 Since the rotating frequency of color wheel is equal to the camera frame rate, which is upper limited to hundreds Hz, the noise introduced by such low speed rotation and vibration  is negligible, as widely used in a commercial projector. In our implementation, we set the camera frame rate as 24 Hz, which can handle daily moving scenes and  a higher frame rate camera can be used for faster movements.

To test the reconstruction accuracy of the hyperspectral images, we use a static color scene to evaluate our imaging system quantitatively. The test scene is strip-wisely uniform and we can calibrate the spectrum with a spectrometer. From the results in Fig.~\ref{reconstruction1}, we can see that the reconstructed spectrum of each patch after GAP optimization is of great consistency with the ground truth spectrum. We highlight the distinction after GAP optimization for clearer observation in the right part.

To demonstrate the dynamic acquisition capability of our method, we conduct two hyperspectral reconstructions of dynamic scenes. The first one is composed of two transmissive scenes with regular motions at a constant speed, as shown in Fig.~\ref{reconstructionvedio}. The translation scene is a color film of Santorini Island mounted on a translation stage moving at a constant speed of 3.7 mm/s. The gray scale camera works at 24 frames per second. We display three frames  in Fig.~\ref{reconstructionvedio}(a), and the hyperspectral vedio of total 274 frames is available online (\href{https://figshare.com/s/73b2e5ed1f129696b1c4}{Visualization 1}).
The rotation one is  a colorful rotating film [Fig.~\ref{reconstructionvedio}(b)] fixed on a rotary translation stage. The angular speed is 0.023 rad/s. The hyeperspectral video showing the whole process is also available online (197 frames in total, see \href{https://figshare.com/s/cb7349d6cfbe4b8632be}{Visualization 2}).

\begin{figure}
  \centering
  \includegraphics[width=1.0\linewidth]{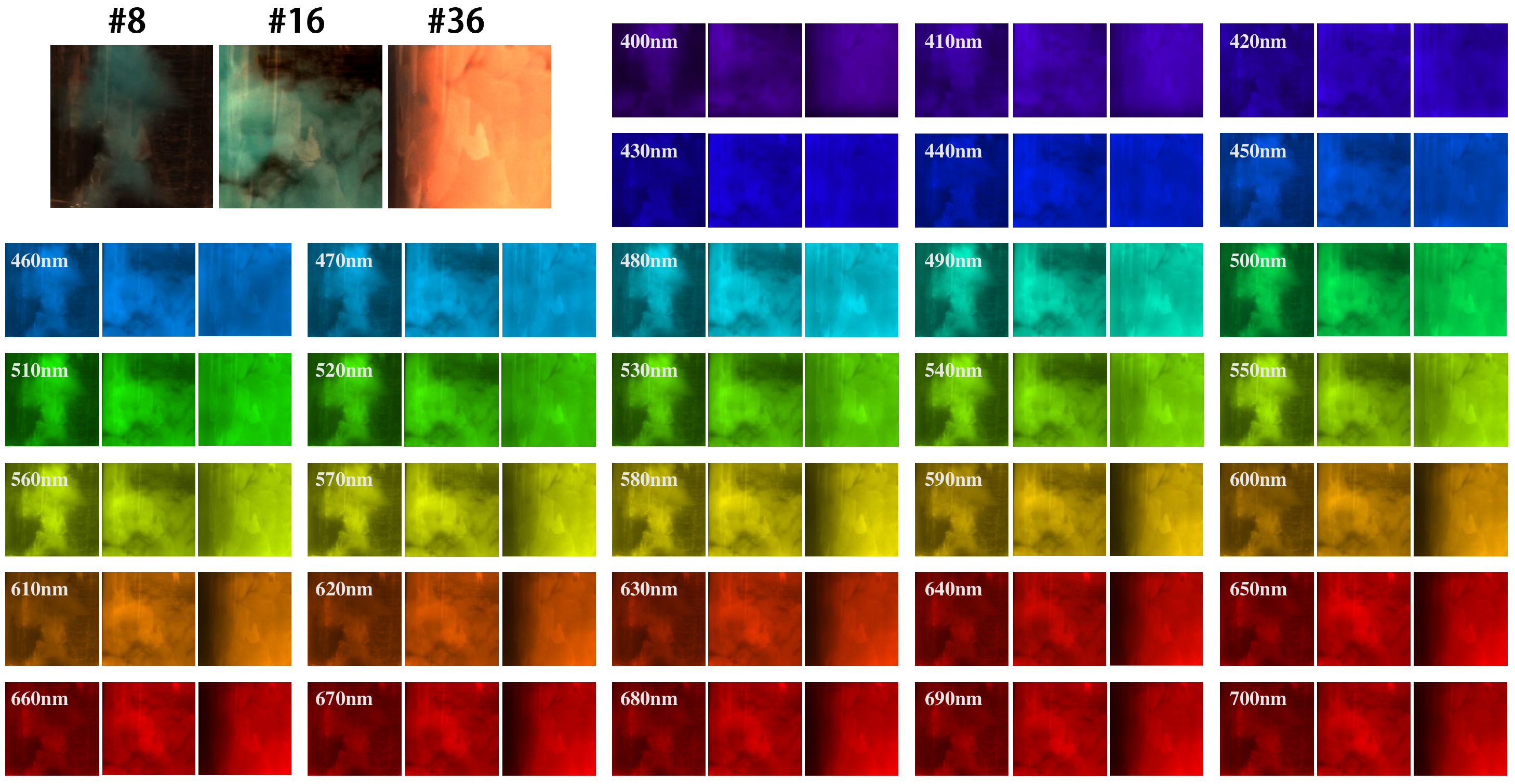}\\
  \caption{Hyperspectral reconstruction of the diffusion process of color pigments poured into clean water. Three out of 77 frames are displayed.}\label{reconstruction3}
\end{figure}

The second experiment is the diffusion process of two color pigments poured into a glass of water, captured in reflective the mode. We first pour the light blue pigment and then the orange one. The color of water gradually changes from light blue to orange, and eventually to the mixture of blue and orange. The spectrum distribution is changing over the diffusion process. We can also see the sinking and mixing process of these two color pigments in the water.
We display  three frames  in Fig.~\ref{reconstruction3} to show the dynamics. The video of total 77 frames is available online (\href{https://figshare.com/s/7fddfa3edeffa58f5781}{Visualization 3}).
From the two reconstructions, we can clearly see that the proposed method can work well for hyperspectral imaging of dynamic scenes.

\section{Summary and discussions}

In summary, we propose a snapshot hyperspectral imaging technique jointly utilizing the spectral and spatial redundancy of hyperspectral data cube. Specifically, we conduct spectral dimension reduction and spatial frequency multiplexing under the computational imaging scheme. For reconstruction, we can resolve spectral transmission/reflectance with low computational load in realtime.
Benefiting from a recently proposed fast sinusoidal modulation \cite{zhang2018simultaneous} working at up to 20 kHz and a rotator working at 1 k revolution per second (rps), our imaging speed is mainly limited by the frame rate of the adopted camera and thus can work well for the common dynamic scenes.
Moreover, utilizing the temporal redundancy of natural scenes,  imaging speed can be further improved by introducing random modulation  based on compressive sensing\cite{hitomi2011video,holloway2012flutter,llull2013coded,reddy2011p2c2,yuan2014low}. The spatial multiplexing scheme would reduce the reconstruction quality slightly in our scheme. To compensate the loss of high frequency details, we can introduce super resolution techniques, either single image based algorithms or sequence based techniques.
 In short, our technique is quite promising to be extended to a high performance hyperspectral imaging system.

It is worth mentioning that the proposed approach might be of sectioning ability for thicker transparent/translucent samples. The sectioning ability depends on the imaging mode: for fluorescent imaging, only the focusing plane is excited and thus we can capture the hyerspectral data of a specific layer; for non-fluorescent imaging, the result would be a summation of different z-planes, including one focusing plane and some out-of-focus ones.
Besides, since we focus on developing a general hyperspectral imaging, six fixed modulation frequencies are used for all the natural scenes. Scene adaptive modulation may be more efficient since samples have different frequency spreads in the Fourier domain, which is worth further studying.

\section*{Funding}
The National key foundation for exploring scientific instrument of China No.2013YQ140517; The National Science Foundation of China under Grant 61327902
and Grant 61631009.



\end{document}